# The CAITLIN Auralization System: Hierarchical Leitmotif Design as a Clue to Program Comprehension

*James L. Alty*
LUTCHI Research Centre
Department of Computer Studies
Loughborough University
Loughborough LE11 3TU, UK
j.l.alty@lboro.ac.uk

*Paul Vickers*
School of Computing and Mathematical Sciences
Liverpool John Moores University
Byrom Street
Liverpool L3 3AF, UK
p.vickers@livjm.ac.uk

**ABSTRACT**
Early experiments have suggested that program auralization can convey information about program structure [8]. Languages like Pascal contain classes of construct that are similar in nature allowing hierarchical classification of their features. This taxonomy can be reflected in the design of musical signatures which are used within the CAITLIN program auralization system. Experiments using these hierarchical leitmotifs indicate whether or not their similarities can be put to good use in communicating information about program structure and state. (Note, at time of going to press experimental results could not be included. These will be presented at the conference and included later.)

**INTRODUCTION**
The CAITLIN pre-processor[1] [8] provides musical auralizations of Turbo® Pascal programs. The system adds calls to library routines to a copy of a program source. These routines generate the musical auralizations which are played via MIDI on a Boss DS-330 multi-timbral synthesiser. Some arguments for using musical techniques have been discussed more elsewhere [1, 2, 8], but in summary:

- Western musical structures, whether by independent evolution or by cultural imposition [7], are widely accepted across the world.

- The information contained in a large scale musical work approaches that of a moving video (a typical audio CD contains hundreds of megabytes). The potential exists for using music to successfully transmit complex information.

- Music forms a large part of peoples' daily lives. It can be very memorable and durable. Most people are reasonably familiar with the language of music in their own culture.

- Music involves the simultaneous transmission of a set of complex ideas related over time, within an established semantic framework. Often, the job of a composer is to use musical resources and techniques to enable a listener to successfully disambiguate such information (although sometimes composers intentionally introduce ambiguity to add interest and depth to their music).

- Music may offer an important communication channel for blind or partially-sighted users.

Our perception of music is primarily temporal, that is, we more readily perceive those features between which there are temporal relationships [11]. It has been observed that when listening to music we tend to perceive it not as an arbitrary sequence of note durations but as a temporal structure in which notes are grouped into various kinds of units [6]. Two features of temporal musical structure, *succession* and *overlap* [11] have analogues in the program domain: sequence and construct nesting. Also, as the structure of music (like that of programs) is multi-levelled [11] and given that the events of an executing program occur in a time-ordered framework, it would seem sensible to attempt to map program events to musical ones. For instance, the musical sonata form provides an *exposition*, *development* and *recapitulation* which could map to the initialization, execution-body and finalisation sections of a typical program and its subprograms.

It has been suggested by Hotchkiss and Wampler [4] that music lends itself well to experiencing data and events subjectively. This, they claim, would give us a greater sense of participation or of being inside a function than is possible using more objective numerical representations. They rather boldly suggest that the sound of an executing program is an "*interesting example of the symphonies that may underlie the running of virtually every computer code.*"

---

[1] http://www.cms.livjm.ac.uk/www/homepage/cmspvick/caitlin/caitlin.htm

In the prototype CAITLIN system, auralizations are effected at the construct level. That is, only the major Pascal constructs (loops and selections) are assigned musical representations. The ultimate aim of the project is to construct a musical debugging environment to assist novice programmers. But before this can be done it must be determined whether music is a useful communication medium in program comprehension. For the purposes of the current work the focus has been deliberately constrained to include only the major constructs and structural features of the computer program. Even with this limitation a measure of success has already been achieved in using the auralizations to help locate bugs in short programs [2].

**POINTS OF INTEREST**
CAITLIN uses the notion of the point of interest (POI) in its auralizations. We defined the POI as "*a feature of a construct, the details of which are of interest to the programmer during execution*" [8]. Each POI is represented by a musical device or leitmotif. In music a leitmotif is a recurring theme associated with a particular thought or character. In the first CAITLIN prototype the leitmotifs, whilst musical in structure, were largely arbitrary in their design, the only consideration being to make each one distinct from the others to avoid ambiguity.

As a starting point this approach was successful, but was limited in that a program's entire auralization did not have a unifying musical structure. Indeed, different constructs could be assigned to different musical scales. The effect was akin to using fragments of separate songs with different meters and musical keys as opposed to passages from a single musical piece linked by a common time signature, tempo and key. The purpose of the current work was to develop a more unified approach to leitmotif design to impose a more formal structure on the musical auralizations.

**PREVIOUS RESULTS**
A previous experiment [8] suggested that programmers could follow the execution of simple programs by listening to their auralizations. Furthermore, they were able to describe the structure of the programs from the information presented by the auralizations. However, a number of limitations in the implementation of the auralizations were highlighted which detracted from the overall effectiveness of the system. These included:
- failure to auralize all POIs of a construct which led to ambiguity;
- poorly designed leitmotifs which led to some failures in subjects' identification of constructs;
- arbitrariness of leitmotif design led to poor association and recall in some subjects causing the same leitmotif to be identified as different constructs on different occasions.

**PROGRAM FEATURES AND REDESIGN OF LEITMOTIFS**
Having achieved a measure of success with simplistic musical devices we concentrated on the details of leitmotif design with a view to making the auralizations more integrated and unified.

Programming languages offer the programmer a range of tools for achieving similar ends. Pascal provides three iteration constructs—WHILE, REPEAT and FOR. Each allows iterative execution of code but differs from the others in the way the looping is controlled. So, the three loop constructs are different but share certain characteristics. Similarity is also found in the selection statements. IF, IF…ELSE, CASE and CASE…ELSE provide for selective execution of statements but use different mechanisms to accomplish it.

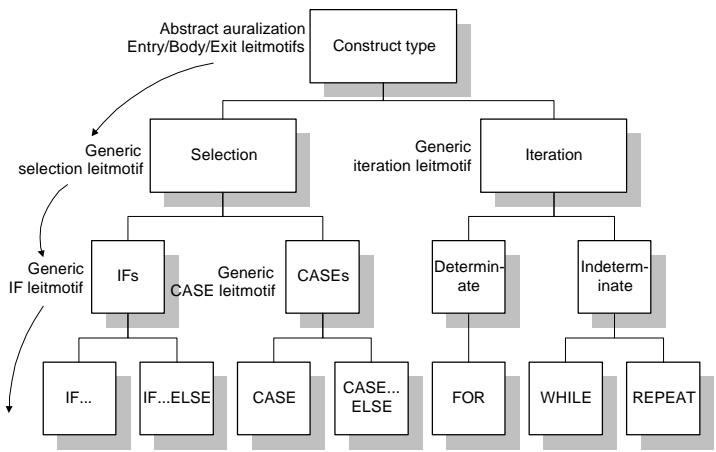

Figure 1: Taxonomy for leitmotif design

So, there is a taxonomy of constructs implicit in the language. The motivation behind the redesigned leitmotifs was to attempt to model this hierarchy musically. In this way there would be a theme denoting iteration and another theme to denote selection. Within each, variations of the theme would be used to represent the individual constructs. All selections would thus sound similar but entirely different from the loop constructs. Figure 1 shows how the selection and iteration constructs of Pascal are organised and how this structure might be modelled in a hierarchical leitmotif design.

The approach taken was to use a chord-based motif for the iterations and a melodic device for the selections. For the two classes of construct there are three fundamental points of interest that require auralization: entry to the construct, execution of the construct's body and exit from the construct. The entry and exit POIs are obviously related as the pair serves to parenthesize the construct. Therefore, these two POIs are modelled by related leitmotifs where the exit motif provides closure for the entry theme. Figure 2 gives the generic selection theme. There is a rising scale signifying entry to the selection construct and a descending scale denoting exit. The underlying theme was reworked for each of the individual selection constructs. This was accomplished by changing the rhythmic patterns of the theme for each leitmotif.

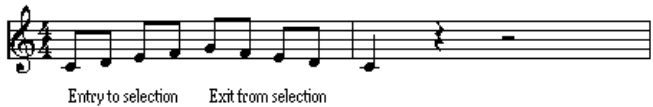

Figure 2: General selection theme

The motif for the simple IF statement (i.e., an IF without an ELSE path) is given as Figure 3. We observe the same basic melodic theme but with a modified rhythm.

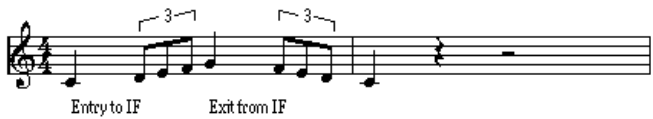

Figure 3: Simple IF statement

The general iteration is given as Figure 4. which shows a simple chord progression of tonic to tonic (I-I) to denote entry to and exit from the loop construct.

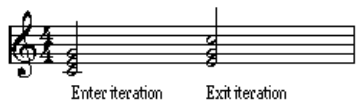

Figure 4: General iteration theme

The FOR loop's variation is shown in Figure 5. We see the overall progression retained but with some additional chords and rhythmical variations. More chords are added to represent the loop bodies (q.v.) leading to quite distinct chord progressions for each loop construct (although each begins and ends on the tonic).

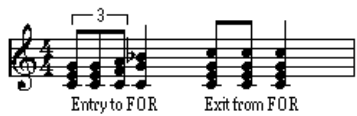

Figure 5: FOR statement theme

In all cases it is important to try to keep the motifs as short as possible while maintaining the ability to generate expectancies [3] of resolution in the mind of the listener. Lessons learned about transition probabilities [9, 11] should assist in the development of more formal guidelines for motif construction.

**CONSTRUCT BODIES**

The principle of construct similarity was used in the design of motifs for the construct bodies. IF and IF…ELSE involve evaluation of Boolean expressions. Iterations use the value of a Boolean expression to control the repeated execution of a group of statements. We represented Boolean evaluations by using a motif in a major key for true results and minor keys for false results. The justification for this seemingly arbitrary choice is that at the top level, the diatonic major and minor modes provide a convenient mapping for the ordinal set of Boolean values (false and true). A potential danger lies in our tendency to associate music in major keys with happiness and the minor modes with sadness. We may subconsciously equate a Boolean true (major) as being 'good' whilst seeing the false as bad (minor). But for the continued execution of the REPEAT loop, the Boolean expression must be false. However, if the major mode is used as the assumed default then this accords with the tendency of adult western listeners to default to a major mode in the absence of information to the contrary [10]. The fact that the REPEAT loop requires a Boolean false (or diatonic minor) to continue its iteration merely serves to highlight the difference in logic between it and the WHILE construct. The mapping of true to major and false to minor has been incorporated in the various points of interest of the constructs. Figure 6 shows the auralization of a simple IF statement whose conditional expression yields true.

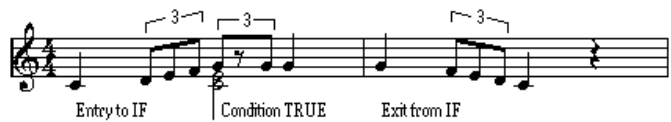

Figure 6: IF statement yielding 'true'

Figure 7 shows the same statement but this time the expression yields false. Notice how the exit motif is also changed to a minor key to reinforce this. Similar devices are used in IF…ELSE and CASE.

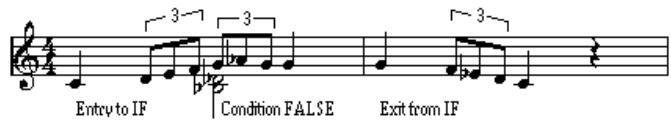

Figure 7: IF statement yielding 'False'

As a loop implies the construct's persistence over time a background drone is added to the iterations' bodies to reinforce in the listener's mind that everything that is happening is doing so *within* a loop. We need to know when the individual loop iterations occur and when the controlling Boolean expression is evaluated. For the REPEAT and WHILE loops a simple major/minor chord device is used when the loop condition is tested; this will be heard immediately after entry to the WHILE loop, but after the iterated statements for the REPEAT loop. Each time one of these chord devices is heard we know that the loop has reached its decision point. The null WHILE loop (where the terminating condition is true upon entry) would thus be heard as a sequence of entry motif followed by minor chord motif for condition evaluation followed by the exit motif.

The FOR loop is count-controlled, a loop invariant taking incremental steps from a starting value to an end value. To denote this stepping up (or down) of the invariant, the pitch of the drone in the FOR loop is increased (or decreased for the FOR…DOWNTO) by one diatonic step with each repetition.

**VARIATIONS ON A THEME**

This idea of variations on a theme ensures that all selections sound like each other but can be distinguished by their individual mutations of the class motif. The reason for doing this is not simply one of organizational convenience, although such categorization can be useful cognitively; applying these techniques provides us with a means of program comprehension at different levels of abstraction and also with a way of conveying spatial information temporally. The abstraction is achieved because one can choose to listen to a program's auralization in terms of its overall structure (e.g. a selection followed by a loop etc.) or in terms of its details (e.g., an IF…ELSE followed by a WHILE). Further abstractions could be achieved by providing selective auralizations in terms of:

- classes of construct to auralize
- number of iterations of a loop
- nesting depth of constructs

**AURAL PRESENTATION OF TEMPORAL AND SPATIAL INFORMATION**

The categorization of constructs also enables a musical portrayal of spatial program features. One of the driving forces behind program auralization is that sound is a temporal medium and program execution is a temporal phenomenon; therefore, it makes sense to explore the possibilities of mapping the latter to the former. Through auralization we can listen to the execution of a program and make inferences about its state. But restricting an auralization to temporal detail alone may lead to a loss of quality.

```
IF a > 3 THEN
  Writeln ('a > 3') ;
IF a > 3 THEN
  Writeln ('a > 3')
ELSE
  Writeln ('a <> 3') ;
```

Figure 8: Anticipation of structure

Consider the code fragment in Figure 8. If all occurrences of IF statements sounded alike and if the value of 'a' were greater than 3 then it would be impossible to determine, from the auralization alone, whether one is hearing a simple selection (first IF in Figure 8) or one with an ELSE path but where the ELSE part was not followed (second IF). By categorizing the constructs and building this into the theme tunes we can avoid this ambiguity. CAITLIN uses a modified form of the simple IF statement motif (Figure 3) to represent the IF statement that has an ELSE part (see Figure 9).

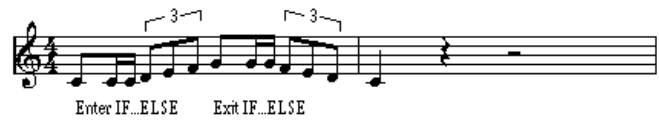

Figure 9: IF…ELSE motif

An advantage of this approach is that when an IF…ELSE occurs and the initial condition is false the listener is not caught unawares by the subsequent occurrence of an ELSE structure in the auralization. By setting up this anticipation[2] by the listener of possible future events CAITLIN creates a sort of construct 'footprint' which shows not just where we have been but also where we might go. Such an auralization is able to capture the spatial information relating to the presence of an ELSE path; the difference between the two constructs in Figure 8 is readily made apparent in the auralization.

The CASE statement (Figure 10) is another example of how CAITLIN conveys spatial information within the musical framework.

```
CASE x OF
  1..4 : Writeln ('Between 1 and 4') ;
  5..7 : Writeln ('Between 5 and 7') ;
  ELSE   Writeln ('No match') ;
END ;
```

Figure 10: CASE statement

Like IF…ELSE…IF, CASE allows for alternative courses of action depending on a variable's value. Of interest to the programmer is which instance of the CASE labels (if any) produces a match. Unlike the IF…ELSE…IF which carries out its comparisons of the various (nested) conditional expressions sequentially, no such ordering is implied by the CASE. However, it is convenient for us to think of the variable as matching the first, second, third etc. instance of the CASE labels. In Figure 10, if 'x' had the value 3, then we would say that the second label produced a match.

This is a spatial judgement because the second label is lower down the list than the first. This fact is communicated by signalling the presence of each label in turn (by a percussive sound). If a particular label produces a match then a major chord is also sounded. The resultant auralization gives the effect of the computer stepping over each label in turn until the end (or ELSE part) is reached or a match occurs.

---

[2] See Robert Jourdain's '*Music, the Brain and Ecstasy*' for an interesting description of musical anticipation [5].

Another spatial element that can be mapped to sound is construct nesting. Programmers show this visually by indenting the code for each level of nesting. Currently CAITLIN represents nesting depth by increasing the octave of nested components. However, it only takes five or six levels of nesting before the pitch becomes too high to be useful. Other possible mappings include position within the stereo field or using background drones for each construct. There are limitations with each approach and further research will be conducted into how best to map nesting to music.

**EXPERIMENT**

**Note: at time of going to press the experiment is still in progress. Results will be made available at a later date.**

To determine whether this hierarchical design approach is useable a small study is being conducted. The aim is to determine whether having heard an example auralization of the iteration and selection classes the subjects could then assign other auralizations to their correct class type. A similar test to that previously described [8] will be performed to see if the new leitmotif designs improve performance in describing a program's structure from its auralization.

The study comprises six experiments. The aims of these are, respectively:

1. To determine whether the two basic types of construct auralization (iteration and selection) are sufficiently distinct for the average user.
2. To determine whether subjects can identify individual constructs after training.
3. To determine whether subjects can identify the various constructs used in short programs where no construct nesting is employed.
4. To determine whether subjects can identify the various constructs used in short programs where combinations of sequential and nested constructs are used.
5. By repeating test 2, to see whether subjects' identification of constructs improved after the repeated exposure given by the intervening tests.
6. To determine whether choice of musical timbre plays a significant role on subjects' ability to discriminate between auralizations.

The subjects will complete a short questionnaire at the start of the study to determine their level of musical training. In each of the six tests, the auralizations are presented in random order to each individual subject.

**RESULTS**

Results from the tests will be analysed to see if there is any support for the ideas set out in the paper. It is expected that subjects will tend to correctly classify the class of each auralization. We will also examine the results to see if a learning effect is evident over the duration of the experiment. We anticipate an improvement in the accuracy of construct recognition using the new motifs. As ambiguities of the first system have been (we hope) removed, the identification of nesting involving selections should improve [8].

**CONCLUSION**

In theory aspects of program structure can be mapped to music. We expect the experimental results to bear this out[3]. The next stage of the project will involve refining the system in the light of these results followed by a further set of experiments involving novice programmers undertaking debugging problems. Also, thorough investigation of the best way to present construct nesting needs to be undertaken—depth of nesting can theoretically be mapped to octave position, stereo imaging, or even multiple background drones could be used, one for each nested construct. The major difficulty with program nesting is that this is a spatial feature rather than a temporal one and so is harder to represent musically. Music is geared more towards temporal rather than spatial aspects. However, the chord is a spatial structure in music and so perhaps there is scope for its use here.

---

[3] See http://www.cms.livjm.ac.uk/www/homepage/cmspvick/caitlin/caitlin.htm for the latest information on the CAITLIN system. Results of the experiment will be shown here and also in the on-line version of the ICAD proceedings.